\documentclass[twocolumn,showpacs,preprintnumbers,amsmath,amssymb]{revtex4}

\usepackage{graphicx}

\begin{document}
\preprint{RUP-08-1}

\title{Decay of the cosmological constant by Hawking radiation as quantum tunneling}
\author{Y.~Sekiwa}
\email{sekiwa@stu.rikkyo.ne.jp}
\affiliation{Department of Physics, Rikkyo University, Tokyo 171-8501, Japan}

\date{February 22, 2008}

\begin{abstract}
We calculate the emission rate of Hawking radiation from the cosmological horizon by quantum tunneling approaches. Using the Hamilton-Jacobi and the null geodesic methods, two typical observations are obtained. First, the spectrum of radiation is not strictly pure thermal. Second, the value of the cosmological constant decreases. Our calculation is different from other similar works about Hawking radiation from the cosmological horizon in that the horizon does not shrink but expands by taking into account the back-reaction effect. We show that when a positive frequency particle is materialized near the cosmological horizon, the contribution of a negative frequency partner to the background geometry lowers the value of the cosmological constant. We also touch upon thermodynamics of de Sitter space and the quantum creation of universe.
\end{abstract}


\maketitle
\section{Introduction}\label{section1}
Understanding of black hole physics is believed to be a hint for the quantum theory of gravity. Among others, Hawking radiation from black holes is one of the most striking effects that is known to arise from the combination of quantum mechanics and general relativity. There are several derivations of Hawking radiation. Hawking's original derivation, which calculates the Bogoliubov coefficients between in and out states for a body collapsing to a black hole, is the most direct and physical \cite{Hawking:1974rv, Hawking:1974sw}. Derivations based on Euclidean quantum gravity are very straightforward \cite{Gibbons:1976ue}. Path integral derivation by Hartle and Hawking uses the Kruskal extension of spacetime and the physical time is analytically continued to the imaginary time. Then Hawking temperature of the spacetime is equal to the imaginary time period of a semiclassical propagator \cite{Hartle:1976tp}. These derivations, however, treat the background geometry fixed, and energy conservation is not enforced during the emission process. Evidently, a black hole loses its mass when particles are emitted. Hence, conventional approaches to derive Hawking radiation can not describe more realistic mass decreasing phenomena.

Recently, Parikh and Wilczek derived Hawking radiation as a quantum tunneling process where energy conservation plays a fundamental role \cite{Parikh:1999mf}. Their tunneling picture is based on a pair creation of particles near the event horizon with zero total energy. If a negative energy particle is closer to the center of the black hole than a positive energy particle, the positive energy particle will see, by virtue of Birkhoff's theorem, as the horizon move inward and may find itself on the outside of the horizon. Then, if the positive energy particle is radiated away from near the outside of the contracted horizon to infinity, the mass of the black hole decreases because the negative energy partner remains inside the horizon of black hole. As a consequence, the horizon contracts from its original radius to a new smaller radius. In this process, the classical one particle action becomes complex and gives the WKB amplitude an imaginary part. The emission rate $\Gamma$ of Parikh and Wilczek provides the rate with a leading order correction arising from the loss of black hole mass which corresponds to the energy carried by the radiated quantum. The result is written down simply as
\begin{equation}
\Gamma\propto\exp\left(\Delta S_{BH}\right) , \label{eq1}
\end{equation}
where $\Delta S_{BH}$ is the change of Bekenstein-Hawking entropy of the black hole between before and after the particle emission. For the low energy limit, eq.(\ref{eq1}) becomes $\Gamma\propto\exp\left(-\beta E\right)$, where $\beta$ is the inverse Hawking temperature and $E$ is the energy of the emitted particle. This is the usual Boltzmann factor. These facts indicate that the spectrum of black hole radiation is not exactly pure thermal. We call this method to calculate the rate $\Gamma$ the null geodesic method. There have been considerable efforts to generalize this work to those of various black holes and the similar results which support Parikh and Wilczek's tunneling picture have been derived \cite{Vagenas:2000am,Medved:2001ca,Vagenas:2001rm,Wu:2006pz,Zhang:2005xt,Zhang:2005wn,Jiang:2005ba,Wang:2007zzm,Hemming:2000as,Vagenas:2002hs,Fang:2005xf,Ali:2007sh,Parikh:2002qh,Medved:2002zj,Zhang:2005sf,Jiang:2005xb,Wu:2006nj,Li:2007kga}.

The other approach to the tunneling from a black hole is the Hamilton-Jacobi method considered by Angheben et al.\cite{Angheben:2005rm}. In this method, the action of a emitted particle satisfies the Hamilton-Jacobi equation. Solving this equation, one finds that the action has an imaginary part. This imaginary part arises from a classically forbidden path for the emitted particle. Due to vacuum fluctuations near the horizon, a pair of particles can be created inside the horizon. If the positive energy particle tunnels out the horizon to infinity, it is observed as a Hawking radiation by a distant observer. In order to assure this process, it is required that a positive energy particle moves across the horizon from inside to outside. This path is obviously a classically forbidden path. This process, however, can be interpreted as a tunneling process quantum mechanically. We call this method the Hamilton-Jacobi method. Several investigations along this work have also been done and led to the correct Hawking temperatures \cite{Nadalini:2005xp,Kerner:2006vu,Zhao:2006zw,Ren:2007xu,Kerner:2007rr}.

The difference of above two methods lies on the part where the classically impossible process occurs. The null geodesic method treats particles classically and the background geometry dynamically, namely, quantum mechanically. It is assumed that particles follow the geodesics and move classically. This is the reason we call it the null geodesic method. This method has the classical impossibility in the background geometry because the horizon of black hole can not contract classically. On the other hand, the Hamilton-Jacobi method treats the background geometry classically, namely, as a fixed background, and the particle which tunnels out the horizon is treated quantum mechanically. In the Hamilton-Jacobi method, the particle does not follow the geodesic but moves the classically forbidden path, namely, across the horizon path.

In this paper, we study Hawking radiation from the cosmological horizon by using the Hamilton-Jacobi and the null geodesic methods. This issue has been studied by using the former method in ref.\cite{Angheben:2005rm}. But their result seems to have the wrong sign. The Hawking temperature with respect to the cosmological horizon has been given by eq.(4.8) of ref.\cite{Angheben:2005rm}. In the vanishing black hole mass limit or empty de Sitter space limit, $r_C\to l$, their inverse Hawking temperature $\beta_C$ reduces to $-2\pi l$, which is obviously negative value. Since negative temperature seems to be unphysical, probably something may be wrong. In ref.\cite{Angheben:2005rm}, the detailed calculation is not given. Thus, we shall reconsider the Hawking temperature of the cosmological horizon by using the Hamilton-Jacobi method. This is the first motivation of our paper.

The calculation of the emission rate from the cosmological horizon by the null geodesic method also has been studied in refs.\cite{Parikh:2002qh,Medved:2002zj} for empty de Sitter space. At first sight, one may think that their calculation is consistent and correct. Although they have considered the process in which the cosmological horizon shrinks by taking into account the particle's self-gravitation, this contradicts the original idea of the null geodesic method. In this method, it is assumed that the positive energy particle follows its geodesic and moves the classically allowed path. If the cosmological horizon shrinks, the particle must move a classically forbidden path in order to materialize as Hawking radiation. In this paper, we consider the different tunneling process from the cosmological horizon using the null geodesic method and derive the new result which is different from refs.\cite{Parikh:2002qh,Medved:2002zj}. This is the second motivation of our paper. In the null geodesic method, we consider the tunneling process that the cosmological horizon expands and the positive frequency particle moves along the classically allowed path. Although our tunneling process is different from that of refs.\cite{Parikh:2002qh,Medved:2002zj}, it matches with the original idea of Parikh and Wilczek's tunneling picture \cite{Parikh:1999mf}. Our result is consistent with the underlying unitary theory, that is, the emission rate is written down in the form of eq.(\ref{eq1}). Furthermore, our result indicates that the value of the cosmological constant decreases due to energy conservation \cite{Sekiwa:2006qj}.

The organization of this paper is as follows. In section \ref{section2}, we calculate the emission rate from the cosmological horizon by using the Hamilton-Jacobi method. Although our result appears to be the unfamiliar expression, eq.(\ref{eq15}), it is interpreted as a usual Boltzmann factor. Section \ref{section3} is devoted to the explanation of the difference between our tunneling picture and that of refs.\cite{Parikh:2002qh,Medved:2002zj}. The emission rate by the null geodesic method is calculated in section \ref{section4}. Finally, in section \ref{section5}, we discuss our result of this paper.

Throughout this paper, the metric signature adopted is $(-, +, +, +)$. The use has been made of natural units, namely $\hbar=c=G=1$ as well as $k=1$.
\section{Hamilton-Jacobi method}\label{section2}
In this section, we calculate the emission rate from the cosmological horizon by using the Hamilton-Jacobi method. In order to explain its derivation in detail, we compare the calculation of the emission rate from the cosmological horizon with that of Schwarzschild black hole. It is shown that the emission rate from the cosmological horizon does not have a familiar form but it is interpreted as a usual Boltzmann factor.

We consider a static spherically symmetric spacetime described by the metric
\begin{equation}
ds^2=-f(r)dt^2+f^{-1}(r)dr^2+r^2d\Omega^2
 , \label{eq2}
\end{equation}
where $d\Omega^2$ is the metric of two dimensional unit sphere. This metric includes the Schwarzschild solution and the de Sitter solution as special cases. In this background, a massless scalar particle follows the Klein-Gordon equation $\nabla_{\mu}\nabla^{\mu}\phi=0$, where $\nabla_{\mu}$ represents the covariant derivative with respect to the metric (\ref{eq2}). We look for the solution having the form $\phi\propto\exp\left(iI+\cdots\right)$, where $I$ denotes the classical action of the particle and dots represents the higher order terms in $\hbar$. If this classical action has an imaginary part, the emission rate $\Gamma$ is given by
\begin{equation}
\Gamma\propto\exp\left(-2\,{\rm Im}\,I\right) . \label{eq3}
\end{equation}
In the following, we calculate the emission rate along the similar line on ref.\cite{Mitra:2006qa} and then we check our result by using the proper distance approach which has been considered originally in ref.\cite{Angheben:2005rm}. Note that in ref.\cite{Mitra:2006qa} the scalar particle is defined as $\phi\propto\exp\left(-iI+\cdots\right)$ so that some signs are different from ours.

Inserting $\phi\propto\exp\left(iI+\cdots\right)$ into the Klein-Gordon equation, one finds the following equation to leading order in $\hbar$,
\begin{equation}
g^{\mu\nu}\partial_{\mu}I\partial_{\nu}I=0 . \label{eq4}
\end{equation}
This is the relativistic Hamilton-Jacobi equation for the classical action of a massless particle in curved spacetime. By the symmetry of the background spacetime, we can use the separation of variables as follows,
\begin{equation}
I=-\omega t+W(r) , \label{eq5}
\end{equation}
where we have ignored the angular part because we consider the s-wave only in this paper. $\omega$ is a constant which corresponds to the frequency of the particle. From eqs.(\ref{eq4}) and (\ref{eq5}), the equation for $W(r)$ becomes
\begin{equation}
-\frac{\omega^2}{f(r)}+f(r)\left(\frac{\partial W(r)}{\partial r}\right)^2=0 \label{eq6}
\end{equation}
and the solution of this equation is expressed as
\begin{equation}
W(r)=\pm \omega\int\frac{dr}{f(r)} , \label{eq7}
\end{equation}
where upper (lower) sign represents that the particle is the outgoing (incoming) wave.

Now we consider the case of Schwarzschild black hole, in which the function $f(r)$ in the metric (\ref{eq2}) is written as $f(r)=1-r_H/r$. Of course, $r_H=2M$ is the radius of the horizon. For the Schwarzschild case, eq.(\ref{eq7}) becomes
\begin{equation}
W(r)=\pm \omega\left(r+r_H\int\frac{dr}{r-r_H}\right) . \label{eq8}
\end{equation}
There is a pole at $r=r_H$ in the $r$ integral. In order to avoid the pole we use Feynman's prescription and replace $r-r_H$ with $r-r_H-i\epsilon$. This yields
\begin{equation}
W(r)=\pm \omega\left[r+i\pi r_H+r_H\int P\left(\frac{1}{r-r_H}\right)dr\right]+\alpha , \label{eq9}
\end{equation}
where $\alpha$ is an integration constant and $P( )$ denotes the principal value. We are interested in the imaginary part of the classical action. From eqs.(\ref{eq5}) and (\ref{eq9}), they are given by
\begin{equation}
{\rm Im}\,I=\pm \pi\omega r_H+{\rm Im}\,\alpha . \label{eq10}
\end{equation}
Incoming particles can fall behind the horizon along classically permitted trajectories, with a capture cross section of order of the horizon area. Hence, the classical action for incoming particles must be real. This determines the imaginary part of a constant $\alpha$, ${\rm Im}\,\alpha=\pi\omega r_H$. Therefore the emission rate from Schwarzschild black hole is written as
\begin{equation}
\Gamma\propto\exp\left(-8\pi M\omega\right) . \label{eq11}
\end{equation}
This is nothing but the Boltzmann factor with the temperature $T_H=1/8\pi M$ which is the Hawking temperature of the Schwarzschild black hole.

Note that in order to obtain the emission rate we have used the boundary condition for the incoming particle. This is equal to considering the ratio between the particle emission and absorption. It is known that semiclassically this ratio is given by $P(\text{emission})=e^{-\beta E}P(\text{absorption})$, where $\beta$ is the inverse Hawking temperature and $E$ is energy of the emitted particle \cite{Hartle:1976tp}.

Along the similar line, we consider the de Sitter case, namely, the emission rate from the cosmological horizon. For de Sitter spacetime, the function $f(r)$ is given by $f(r)=1-r^2/r_H^2$, where $r_H=l=\sqrt{3/\Lambda}$ is the radius of the cosmological horizon and $\Lambda$ is the cosmological constant. In this case, eq.(\ref{eq7}) has the form
\begin{equation}
W(r)=\mp \omega\frac{r_H}{2}\left[-\log\left(r+r_H\right)+\int\frac{dr}{r-r_H}\right]+\alpha . \label{eq12}
\end{equation}
In order to avoid the pole, we again use Feynman's prescription and replace $r-r_H$ with $r-r_H+i\epsilon$ instead of $r-r_H-i\epsilon$. This is because the positive frequency particle is the incoming wave for the radiation from the cosmological horizon. Although the sign of $i\epsilon$ is reversed in comparison with that of the Schwarzschild case, the positive frequency particle decays in time for both the Schwarzschild and de Sitter cases by these choices. Then, eq.(\ref{eq12}) yields
\begin{equation}
\begin{split}
W(r)=&\mp \omega\frac{r_H}{2}\bigg[-i\pi-\log\left(r+r_H\right) \\
&+\int P\left(\frac{1}{r-r_H}\right)dr\bigg]+\alpha , 
\end{split}\label{eq13}
\end{equation}
and the imaginary part of the classical action becomes
\begin{equation}
{\rm Im}\,I=\pm\frac{\pi}{2}\omega r_H+{\rm Im}\,\alpha . \label{eq14}
\end{equation}
For the de Sitter case, since outgoing particles can move away beyond the horizon along classically allowed paths, their actions are real. This gives ${\rm Im}\,\alpha=-\pi\omega r_H/2$. Therefore the emission rate from the cosmological horizon becomes
\begin{equation}
\Gamma\propto\exp\left(2\pi l\omega\right) . \label{eq15}
\end{equation}
If one identifies $\beta_C=-2\pi l$ as the inverse Hawking temperature of the cosmological horizon, the result of ref.\cite{Angheben:2005rm}, $\Gamma\propto\exp\left(-\beta_C\omega\right)$, is reproduced. Since negative temperature seems to be unphysical, however, we propose a different interpretation of the emission rate (\ref{eq15}). Before doing it, we shall derive the emission rate (\ref{eq15}) by an another approach in which the proper spatial distance is used. As Angheben et al. have used originally the proper spatial distance to calculate the emission rate \cite{Angheben:2005rm}, we check whether the replacement $r-r_H\to r-r_H+i\epsilon$ is consistent with the proper distance approach.

For the metric (\ref{eq2}), the proper distance $\sigma$ is defined by
\begin{equation}
\sigma=\int\frac{dr}{\sqrt{f(r)}} . \label{eq16}
\end{equation}
Using this proper distance, eq.(\ref{eq7}) is written as
\begin{equation}
W(r)=\pm \omega\int\frac{d\sigma}{\sqrt{f(\sigma(r))}} . \label{eq17}
\end{equation}
Since, near the horizon, function $f(r)$ is expanded as $f(r)\simeq f'(r_H)(r-r_H)+\cdots$, the proper distance becomes
\begin{equation}
\sigma\simeq\frac{2}{\sqrt{f'(r_H)}}\sqrt{r-r_H} . \label{eq18}
\end{equation}
Then $W(r)$ has the following form,
\begin{equation}
W(r)=\pm\frac{2\omega}{|f'(r_H)|}\int\frac{d\sigma}{\sigma} . \label{eq19}
\end{equation}

First, we shall consider the Schwarzschild case where $|f'(r_H)|=1/2M$. To avoid the pole at $\sigma=0$, we use Feynman's prescription $\sigma \to \sigma-i\epsilon$. Then the imaginary part of the classical action is given by
\begin{equation}
{\rm Im}\,I=\pm 2\pi M\omega+{\rm Im}\,\alpha , \label{eq20}
\end{equation}
where $\alpha$ is a constant. Here we have used the fact that the integral contour is transformed from a semicircle to a quarter circle by coordinate transformation $r\to\sigma$ \cite{Akhmedov:2006pg} which is given by eq.(\ref{eq18}). This contour yields the factor $i\pi/2$ on the imaginary part rather than $i\pi$. Equation (\ref{eq20}) is the same with eq.(\ref{eq10}) so that the emission rate from Schwarzschild black hole is given by eq.(\ref{eq11}). This is consistent with the previous result for the emission rate.

Next, we consider the de Sitter case. One may expect that if the same calculation with the Schwarzschild case is done, the correct result will be derived. There is, however, the difference between the Schwarzschild case and the de Sitter case with respect to the definition of the proper spatial distance. For the Schwarzschild case, from eq.(\ref{eq16}) the proper distance is given by
\begin{equation}
\sigma=\sqrt{r(r-r_H)}+\frac{r_H}{2}\log\left|\frac{\sqrt{r}+\sqrt{r-r_H}}{\sqrt{r}-\sqrt{r-r_H}}\right| , \label{eq21}
\end{equation}
where we have chosen an integration constant as $\sigma=0$ at $r=r_H$. This proper distance is positive definite as long as $r>r_H$. On the other hand, for the de Sitter case, if the same definition, eq.(\ref{eq16}), is used, one finds
\begin{equation}
\sigma=r_H\left[\arcsin\left(\frac{r}{r_H}\right)-\frac{\pi}{2}\right] . \label{eq22}
\end{equation}
As before, an integration constant has been chosen as $\sigma=0$ at $r=r_H$. This proper distance is negative definite when $r<r_H$. For the Schwarzschild case, the outgoing wave which has the positive frequency propagates toward the $\sigma$ increasing direction. On the other hand, for the de Sitter case, the incoming wave which has the positive frequency propagates toward the $\sigma$ decreasing direction.

To avoid the sign ambiguity of the $i\epsilon$ prescription, we use the same relation between the direction of motion of the positive frequency particle and the direction of $\sigma$ axis for both the Schwarzschild and de Sitter cases. In order to make $\sigma$ positive definite, we define the proper distance as
\begin{equation}
\sigma=-\int\frac{dr}{\sqrt{f(r)}} \label{eq23}
\end{equation}
for the de Sitter case instead of eq.(\ref{eq16}). Then the proper distance becomes
\begin{equation}
\sigma=r_H\left[\frac{\pi}{2}-\arcsin\left(\frac{r}{r_H}\right)\right] . \label{eq24}
\end{equation}
This proper distance is positive definite when $r<r_H$ and measures the proper distance from the cosmological horizon $r_H$ to a certain point $r$ $(r<r_H)$. By this definition, the incoming wave which has the positive frequency moves toward the $\sigma$ increasing direction as the Schwarzschild case.

Note that the minus sign in eq.(\ref{eq23}) arises from the fact that the upper bound and the lower bound of the integral is reversed. For the Schwarzschild case, in eq.(\ref{eq16}), it is implicitly assumed that the upper bound is larger than the lower bound. On the other hand, for the de Sitter case, we measure the proper distance from the horizon $r_H$ to a certain point $r$ $(r<r_H)$ so that the upper bound is smaller than the lower bound. To keep the larger radius as the upper bound of the integral, we have inserted the minus sign in eq.(\ref{eq23}).

One can obtain the imaginary part of the classical action from eq.(\ref{eq19}) with the replacement $\sigma\to -\sigma$, but the expression for $W(r)$ remains eq.(\ref{eq19}), where the upper bound is larger than the lower bound of the integral. Next, we use Feynman's prescription in order to avoid the pole. Since we have expressed the action by the proper distance, the positive frequency particle moves toward $\sigma$ increasing direction for both the Schwarzschild and de Sitter cases. So we need not now take into consideration the direction of motion when Feynman's prescription is used. In other words, for both spacetimes, we can use the replacement $\sigma\to\sigma-i\epsilon$ equally. Consequently, the imaginary part of the action becomes
\begin{equation}
{\rm Im}\,I=\pm \frac{\pi}{2}\omega r_H+{\rm Im}\,\alpha , \label{eq25}
\end{equation}
where $\alpha$ is a constant. This is the same expression with eq.(\ref{eq14}) so that the emission rate from the cosmological horizon is given by eq.(\ref{eq15}) again.

Finally, we propose a new interpretation of the emission rate (\ref{eq15}). As mentioned above, if one identifies the inverse Hawking temperature of the cosmological horizon as $\beta_C=-2\pi l$, the emission rate is written as
\begin{equation}
\Gamma\propto\exp\left(-\beta_C\omega\right) , \label{eq26}
\end{equation}
which appears to be a usual Boltzmann factor and this agrees with the result of ref.\cite{Angheben:2005rm}. Negative temperature, however, seems to be unphysical for the emission rate. Rather, it is natural to consider that temperature is always positive. In this paper we stand this point of view and identify $\beta_C=2\pi l$ as the inverse Hawking temperature. Then the Hawking temperature of the cosmological horizon is, of course, given by a well known result $T_H=1/2\pi l$ \cite{Gibbons:1977mu}. Using this temperature, the emission rate (\ref{eq15}) becomes $\Gamma\propto\exp\left(\beta_C\omega\right)$. We write it as follows,
\begin{equation}
\Gamma\propto\exp\left(-\beta_CE\right) \quad ; \quad E=-\omega . \label{eq28}
\end{equation}

In order to explain the physical meaning of eq.(\ref{eq28}), we shall recall the Schwarzschild-de Sitter solution which is represented by the metric (\ref{eq2}) with
\begin{equation}
f(r)=1-\frac{2m}{r}-\frac{r^2}{l^2} . \label{eq29}
\end{equation}
Hereafter we distinguish the terms ``energy'' and ``mass''. The Schwarzschild-de Sitter solution approaches the usual Schwarzschild solution in the limit $l\to\infty$ so that the parameter $m$ represents the mass of black hole. On the other hand, it is known that the energy of this system is given by $E=-m$ \cite{Balasubramanian:2001nb}. That is, positive ``mass'' is measured as negative ``energy'' in de Sitter space.

Similarly, we shall distinguish the terms ``energy'' and ``frequency'' when we discuss the problem of the particle emission from the cosmological horizon in empty de Sitter space. If a emitted particle has positive ``frequency'' $\omega$, it will be measured as negative ``energy'' $E=-\omega$. If this particle collapses to a black hole near $r=0$, then the metric after the collapse will be described by eqs.(\ref{eq2}) and (\ref{eq29}) with $m=\omega$. This spacetime has the black hole mass $\omega$ and the energy $E=-\omega$. It is reasonable to consider that this collapse will occur naturally because the total energy is conserved before and after the collapse. From these facts, we can recognize that for a particle with the positive frequency $\omega$, its energy is expressed as $E=-\omega$. Then the emission rate is written by means of this energy as eq.(\ref{eq28}). As a result, we find that eq.(\ref{eq15}) or (\ref{eq28}) represents the usual Boltzmann factor with the inverse Hawking temperature $\beta_C=2\pi l$.
\section{Null geodesic method}\label{section3}
In this section, we consider the particle emission by the null geodesic method. Here we stress the fundamental difference between our approach and others. The emission rate is calculated concretely in the next section.

To begin with, we review the original idea of quantum tunneling by Parikh and Wilczek \cite{Parikh:1999mf}. Their tunneling picture is based on a pair creation of particles near the horizon with zero total energy. If a negative energy particle is closer to the center of the black hole than a positive energy particle, the positive energy partner will see the horizon shrink and may find itself on the outside of the black hole. The contribution of the negative energy particle to the background geometry makes the event horizon contract from the original radius to the new smaller one. If the positive energy particle lies on the outside of the new horizon and propagates classically to infinity, it is observed as Hawking radiation.

To describe across-horizon phenomena, it is necessary to choose coordinates which are not singular at the horizon. A particular suitable choice is Painlev\'e coordinates. For Schwarzschild black hole, it is written as follows \cite{Parikh:1999mf},
\begin{equation}
ds^2=-\left(1-\frac{2M}{r}\right)dt^2+2\sqrt{\frac{2M}{r}}dtdr+dr^2+r^2d\Omega^2 . \label{eq30}
\end{equation}
Then the radial null geodesics are given by
\begin{equation}
\dot{r}=\pm 1-\sqrt{\frac{2M}{r}} , \label{eq31}
\end{equation}
with the upper (lower) sign corresponding to the outgoing (incoming) geodesics, under the implicit assumption that $t$ increases toward the future. These equations are modified when the particle's self-gravitation is taken into account. If the total energy is fixed by energy conservation, we should use eqs.(\ref{eq30}) and (\ref{eq31}) with $M\to M-\omega$, where $\omega$ is the energy of the emitted particle.

\begin{figure}[tb]
\begin{center}
\includegraphics{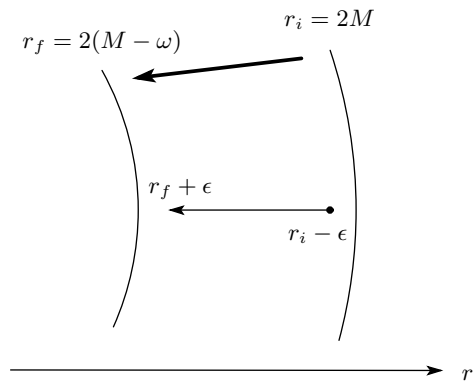}
\end{center}
\caption{Tunneling picture from Schwarzschild black hole. The event horizon contracts from the initial radius $r_i=2M$ to the final radius $r_f=2(M-\omega)$. A pair creation occurs at $r_i-\epsilon$ from which the {\it outgoing} positive energy particle starts to move along its geodesic. The particle materializes as a real particle at $r_f+\epsilon$, from which it propagates classically to infinity.}
\label{fig1}
\end{figure}

The null geodesic method assumes that the positive energy particle follows the classical orbit, namely, its geodesic. When a pair creation occurs inside the horizon, one finds, from eq.(\ref{eq31}), both incoming and outgoing particles move toward $r=0$. Hence, particles can not escape from the black hole. In the idea of Parikh and Wilczek, one treats the background geometry dynamically. As mentioned above, the contribution of the negative energy particle makes the horizon contract. This is the classically impossible process. Classically, black hole can increase the area according to Hawking's area theorem, but can not decrease it or radius. In the null geodesic method, this impossibility is not valid due to quantum mechanical effect by which the particle emission can occur. By energy conservation, if the particle which has the positive energy $\omega$ is emitted, the radius of black hole must contract. The authors of ref.\cite{Parikh:1999mf} express these phenomena in their paper as follows. ``Although this radially inward motion appears at first sight to be classically allowed, it is nevertheless a classically forbidden trajectory because the apparent horizon is itself contracting'' and ``the outgoing particle starts from $r=2M-\epsilon$, just inside the {\it initial} position of the horizon, and traverses the contracting horizon to materialize at $r=2(M-\omega)+\epsilon$, just outside the {\it final} position of the horizon''. This situation is drawn in Fig.\ref{fig1}. There are two key points in the null geodesic method. First, the positive energy particle moves along a classically allowed path or its geodesic. Second, the contracting horizon passes the positive energy particle from behind. Therefore the radius of the initial horizon must be larger than that of the final horizon of the black hole.

Along the similar line, we consider the tunneling from the cosmological horizon. The line element of de Sitter space in Painlev\'e coordinates is expressed as follows \cite{Parikh:2002qh},
\begin{equation}
ds^2=-\left(1-\frac{r^2}{l^2}\right)dt^2-2\frac{r}{l}dtdr+dr^2+r^2d\Omega . \label{eq32}
\end{equation}
The radial null geodesics are given by
\begin{equation}
\dot{r}=\frac{r}{l}\pm 1 , \label{eq33}
\end{equation}
where upper (lower) sign represents the outgoing (incoming) geodesics. When the particle is beyond the horizon $(r>l)$, both outgoing and incoming geodesics correspond to increasing $r$ and the particle never crosses the horizon classically. This is the important fact in order to apply the null geodesic method to the cosmological horizon.

\begin{figure}[tb]
\begin{center}
\includegraphics{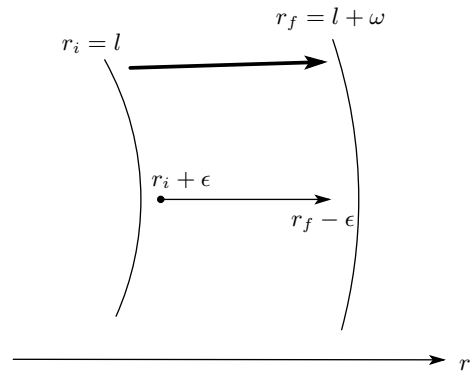}
\end{center}
\caption{Tunneling picture from the cosmological horizon. The horizon expands from the initial radius $r_i=l$ to the final radius $r_f=l+\omega$ (see section \ref{section4} for the details of the final radius $r_f$). A pair creation occurs at $r_i+\epsilon$ from which the {\it incoming} positive frequency particle starts to move along its geodesic. The particle materializes as a real particle at $r_f-\epsilon$, from which it propagates classically inward the origin ($r=0$).}
\label{fig2}
\end{figure}

Suppose that the pair creation occurs outside the horizon. If the incoming particle starts from $r=l+\epsilon$, just outside the initial position of the horizon, it moves toward $\dot{r}>0$ direction according to its geodesic. In order to materialize at the point just the final position of the horizon, the cosmological horizon must expand. So the phenomenon drawn in Fig.\ref{fig2} is required to emit the particle from the cosmological horizon. Hence the contribution of the negative frequency particle to the background geometry must make the cosmological horizon expand.

\begin{figure}[tb]
\begin{center}
\includegraphics{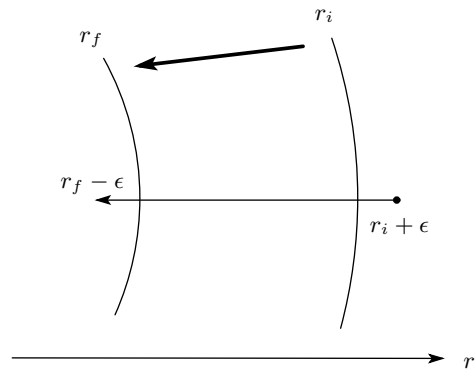}
\end{center}
\caption{Tunneling picture which has been considered in refs.\cite{Parikh:2002qh, Medved:2002zj}. The cosmological horizon contracts from the initial radius $r_i$ to the final radius $r_f$, because the background changes from empty de Sitter space to Schwarzschild-de Sitter space. A pair creation occurs at $r_i+\epsilon$. The path from $r_i+\epsilon$ to $r_f-\epsilon$ is the classically forbidden one. Thus, it is not allowed to use this path in the null geodesic method.}
\label{fig3}
\end{figure}

Meanwhile, the quantum tunneling from the cosmological horizon has been already studied by Parikh \cite{Parikh:2002qh} and Medved \cite{Medved:2002zj} for empty de Sitter space. They, however, have assumed that the background geometry changes from empty de Sitter space to Schwarzschild-de Sitter space. If this happens, the radius of the cosmological horizon contracts $(r_f<r_i)$. In this case, if the positive frequency particle follows its geodesic, the tunneling can not occur, because the horizon does not pass the particle from behind. Both authors of refs.\cite{Parikh:2002qh, Medved:2002zj}, further, have assumed that the pair creation occurs just outside the initial horizon and the positive frequency particle tunnels out just inside the final horizon. Indeed, Parikh claims in ref.\cite{Parikh:2002qh} as follows. ``We expect $r_i$ to correspond roughly to the site of pair creation, which should be slightly outside the horizon'' and ``We expect $r_f$ to be a classical turning point, at which the semiclassical trajectory ({\it i.e.} instanton) can join onto a classical-allowed motion. This must be slightly within the horizon.'' Under the assumption that the cosmological horizon shrinks, this situation is drawn as Fig.\ref{fig3}. In this process, obviously, the emitted particle does not follow the classically allowed path because all particles move toward $\dot{r}>0$ direction for $r>l$.

From these facts, we think that the calculation of the tunneling rate from the cosmological horizon studied in refs.\cite{Parikh:2002qh, Medved:2002zj} may be wrong or does not match with the original idea of quantum tunneling constructed in ref.\cite{Parikh:1999mf}. By these reasons, in the next section, we investigate the emission rate from the cosmological horizon using the null geodesic method based on Fig.\ref{fig2}. The fundamental difference between ours and refs.\cite{Parikh:2002qh, Medved:2002zj} lies on the change of the horizon radius and the path of the positive frequency particle.
\section{The emission rate}\label{section4}
In this section, we calculate the emission rate from the cosmological horizon based on the consideration of previous section and Fig.\ref{fig2}.

The imaginary part of the action of an s-wave incoming particle which crosses the horizon can be written as \cite{Parikh:1999mf}
\begin{equation}
{\rm Im}\, I={\rm Im}\int _{r_i}^{r_f}p_r={\rm Im}\int_{r_i}^{r_f}\int_{0}^{p_r}dp_r'dr . \label{eq34} 
\end{equation}
Using Hamilton's equation, we change the variable from momentum to energy, then we have
\begin{equation}
{\rm Im}\, I={\rm Im}\int_{r_i}^{r_f}\int\frac{dH}{\dot{r}}dr . \label{eq35}
\end{equation}

Here we must consider two problems. The first problem is what is the total energy or Hamiltonian of de Sitter space. The other is how to include the back-reaction effect by particle's self-gravitation. We can estimate the variation of Hamiltonian $dH$ easily. In section \ref{section2}, we have given the energy of the emitted positive frequency particle as $E=-\omega$. Since the pair creation occurs with zero total energy, the negative frequency partner should have the energy $E=+\omega$. Thus, the variation of Hamiltonian is given by $dH=+d\omega$. Meanwhile, for the change of the horizon radius, we can also have the simple estimation as follows. First, temperature and entropy of de Sitter space is $T_H=1/2\pi l$ and $S=\pi l^2$. Since de Sitter space is one parameter spacetime as Schwarzschild spacetime, one can deduce that temperature, entropy and energy of the system satisfies the relation
\begin{equation}
dE=T_HdS . \label{eq36}
\end{equation}
For de Sitter space, the right hand side of eq.(\ref{eq36}) yields $T_HdS=dl$. Hence, the total energy of de Sitter space should have the form $E=l+{\rm const.}$, where constant term is not important for latter considerations. Thus we can use $H=l$ as the Hamiltonian. When the contribution of the negative frequency particle to the background geometry is taken into account, the energy $+\omega$ is added and one has $H'=l+\omega$ as the modified Hamiltonian. Therefore, the effect of back-reaction makes the horizon expand from the original radius $r_i=l$ to the new larger one $r_f=l+\omega$. This is the expected result, because the expansion of the radius of the cosmological horizon does not contradict with the tunneling picture of Fig.\ref{fig2}.

Although the fact that the contribution of the negative frequency particle increases the energy of the system appears at first sight to be curious, this can be understood naturally. The horizon radius $l$ is related to the cosmological constant $\Lambda$ as $l=\sqrt{3/\Lambda}$. When $l$ becomes large, $\Lambda$ must decrease. Then the vacuum energy density $\Lambda/8\pi$ also decreases. Thus, one finds intuitively that the true energy of the system never increases. The origin of the reversed sign for the energy lies on the negative binding energy between matters and gravitational field of de Sitter space \cite{Myung:2001ab}.

Returning to the calculation of the emission rate, we use $dH=+d\omega$ and the replacement $l\to l+\omega$. Then the imaginary part of the action becomes
\begin{equation}
{\rm Im}\, I={\rm Im}\,\int_{r_i}^{r_f}\int_{0}^{\omega}\frac{l+\omega'}{r-(l+\omega')+i\epsilon}drd\omega' , \label{eq37}
\end{equation}
where we have used Feynman's prescription, $\omega'\to \omega'-i\epsilon$. Performing the integrations, eq.(\ref{eq37}) results in
\begin{equation}
{\rm Im}\, I=-\pi\left(l\omega+\frac{1}{2}\omega^2\right) , \label{eq38}
\end{equation}
then the emission rate has the following form,
\begin{equation}
\Gamma\propto\exp\left(2\pi l\omega+\pi\omega^2\right) . \label{eq39}
\end{equation}
We find that the leading term gives the thermal Boltzmann factor for the radiation, and that the second term represents the correction from the response of the background geometry to the emission of a quantum. In fact, the leading term agrees with eq.(\ref{eq15}). Hence, our interpretation about the energy in section \ref{section2} is consistent with the result by the null geodesic method. That is, the leading emission rate is written by means of $\beta_C=2\pi l$ and $E=-\omega$ as eq.(\ref{eq28}).

If the entropy expression is used, equation (\ref{eq39}) can be written in the more elegant form. Since the initial entropy of de Sitter space is $S_i=\pi l^2$ and the final entropy is $S_f=\pi(l+\omega)^2$, one can write the emission rate, eq.(\ref{eq39}), as follows,
\begin{equation}
\Gamma\propto\exp\left(S_f-S_i\right)=\exp\left(\Delta S_{BH}\right) . \label{eq40}
\end{equation}
This expression agrees with eq.(\ref{eq1}) which is the important result of ref.\cite{Parikh:1999mf}. So, the emission rate is represented by the entropy difference between before and after the particle emission.

Although we do not have the exact expression of the emission rate including the quantized gravity, quantum field theory tells us that the rate must be written as \cite{Parikh:2002qh}
\begin{equation}
\Gamma\propto|\mathcal{M}_{fi}|^2\times\text{ (phase space factor) } , \label{eq41}
\end{equation}
where $\mathcal{M}_{fi}$ is the amplitude for the emission process. The phase space factor is obtained by summing over the final states and averaging over the initial states. Since the number of states is just the exponent of the entropy, the phase space factor is given by $e^{S_f}/e^{S_i}=\exp\left(\Delta S_{BH}\right)$. This is nothing but eq.(\ref{eq1}) or (\ref{eq40}) and agrees with the our result. Therefore our result shows that the emission rate deviates from the pure thermal emission and it is in full consistency with the unitary theory.

Finally, we note that our consideration of the tunneling from the cosmological horizon have used the null geodesic method. In this method, since the positive frequency particle follows its geodesic, the cosmological horizon must expand or the final radius of the cosmological horizon must become larger than the initial radius. Suppose that the initial horizon is $r_i=l_i$ and the final horizon $r_f=l_f$. Then, since the horizon radius is related to the cosmological constant $\Lambda$ as $l=\sqrt{3/\Lambda}$, the relation $r_i<r_f$ leads us to the result
\begin{equation}
\Lambda_f < \Lambda_i , \label{eq42}
\end{equation}
where $\Lambda_i$ is the initial cosmological constant and $\Lambda_f$ is the final one. This means that the value of the cosmological constant decreases by the particle emission from the cosmological horizon. This will afford an active credence to our previous work on the thermodynamic consideration of the cosmological constant \cite{Sekiwa:2006qj}.

Our consideration can be applied to not only empty de Sitter solution but also black hole solutions in de Sitter space which have two event horizons. Some investigations similar to refs.\cite{Parikh:2002qh, Medved:2002zj} have been done for these solutions \cite{Zhang:2005sf,Jiang:2005xb,Wu:2006nj,Li:2007kga}, in which they have considered the process that the cosmological horizon contracts. Thus, their results may not describe the tunneling process by the null geodesic method on the correct way. For the cosmological horizon of black hole solutions in de Sitter space, similaly, one needs to consider the horizon expanding process in order to assure quantum tunneling.
\section{Discussion}\label{section5}
In this paper, using the Hamilton-Jacobi and the null geodesic methods, we have investigated the emission rate from the cosmological horizon by quantum tunneling. We have shown that the leading order emission rate is given by eq.(\ref{eq15}) or (\ref{eq28}) and that the rate which includes the response of the background geometry is given by eq.(\ref{eq39}) or (\ref{eq40}).

There are two important key points in our consideration. The first is the interpretation of the emission rate by the sign reversed energy, eq.(\ref{eq28}). In de Sitter space, the energy of a black hole with mass $m$ is evaluated as $E=-m$ \cite{Balasubramanian:2001nb}. Along the similar line, we have proposed that the emitted particle should have energy $E=-\omega$, where $\omega$ is the positive frequency of the particle. Then temperature of de Sitter space corresponds to the usual result $\beta_C=2\pi l$ \cite{Gibbons:1977mu}. The other key point is the change of the radius of the cosmological horizon in the null geodesic method. When the particle's self-gravitation is taken into account, the horizon should expand in order to assure quantum tunneling. If the horizon contracts, the pair created particle can not tunnel out the horizon. The authors of refs.\cite{Parikh:2002qh, Medved:2002zj} have assumed that by a pair creation the background geometry changes from empty de Sitter space to Schwarzschild-de Sitter space. This, however, leads to the contraction of the cosmological horizon. Although they have considered that the positive frequency particle collapses to a black hole, the unobserved negative frequency particle does not contribute to the background geometry in their process. It contradicts with the original idea of quantum tunneling constructed in ref.\cite{Parikh:1999mf}. From this reason, we have reconsidered the tunneling process in this paper.

Our study has an another important implication about the cosmological constant. When the radius of the cosmological horizon expands, the value of the cosmological constant must decrease through the relation $l=\sqrt{3/\Lambda}$. If $\Lambda$ does not decrease, the cosmological horizon does not expand. Therefore, the pair created negative frequency particle should contribute to the background geometry as the horizon radius expands or the value of the cosmological constant decreases. This will provide a strong confirmation to our previous work about thermodynamic consideration of the cosmological constant \cite{Sekiwa:2006qj}.

The decrease of the cosmological constant has physical meaning as follows. First, the second law of thermodynamics holds for de Sitter space. Since the entropy of de Sitter space is given by $S=3\pi/\Lambda$, $S$ increases according to the decrease of $\Lambda$. Furthermore, the third law of thermodynamics also holds for de Sitter space. Temperature of de Sitter space is given by $T_H=1/2\pi l$ or $T_H=(1/2\pi)\sqrt{\Lambda/3}$. As quantum tunneling occurs, $\Lambda$ continues to decrease but never be zero or negative. This is because the tunneling never happens if the final horizon does not exist. In order to exist the cosmological horizon, the cosmological constant $\Lambda$ must be positive. Therefore, de Sitter space remains to be de Sitter space decreasing $\Lambda$ gradually. According to the third law of thermodynamics, temperature of the system never reaches to zero by finite physical processes so that the cosmological constant never has the zero value. Even if $\Lambda$ approaches zero, it never takes the exact zero value. Second, the Painlev\'e coordinates (\ref{eq32}) and planar coordinates
\begin{equation}
ds^2=-dt^2+e^{2t/l}(d\rho^2+\rho^2d\Omega^2)
\end{equation}
are related each other by the coordinate transformation $r=\rho\,e^{t/l}$ so that the time coordinates are the same in both coordinates. Thus, the decay rates of the cosmological constant per unit time are also the same. Planar coordinates describe the exponential expansion of our universe with a scale factor $e^{t/l}$. Therefore, our universe will be de Sitter with decreasing $\Lambda$ in future. From this point of view, our result does not seem to contradict with the observational value of the cosmological constant which is extremaly small. Third, if one believes the no boundary proposal for the quantum creation of universe by Hartle and Hawking \cite{Hartle:1983ai}, the most probable state is given by the nearly zero cosmological constant state \cite{Hawking:1984hk, Wu:2007ht}. The decrease of the cosmological constant by quantum tunneling may imply that our universe gradually makes a transition towards the most probable state, that is, $\Lambda\to 0$ state. Our result indicates that semiclassical consideration does not contradict with these early works. 

The thermal decay of the cosmological constant has been studied recently in refs.\cite{Gomberoff:2003zh,Gomberoff:2005je}. They have treated the decay process as quantum tunneling or thermal activation. It is the natural next step to consider the relation between their decay process of the cosmological constant and our tunneling approach for the cosmological horizon. Although we do not still have the detailed mechanism for the decay of the cosmological constant by Hawking radiation from the cosmological horizon, the underlying physics is probably the same. We hope that this tantalizing problem will be understood in the future.

\end{document}